\documentclass[twocolumn,showpacs,preprintnumbers,amsmath,amssymb]{revtex4}

\usepackage{graphicx}
\usepackage{dcolumn}
\usepackage{bm}
\usepackage{amsmath}

\begin{document}
\title{Study of $^{26}$Mg through 1p pick up reaction $^{27}$Al(d,$^{3}$He)}
\author{Vishal Srivastava$^1$\footnote{Email: vis.vip22@gmail.com\\$^{\nparallel}$Raja Ramanna Fellow}, C. Bhattacharya$^1$,
T. K. Rana$^1$, S. Manna$^1$, S. Kundu$^1$, S. Bhattacharya$^{1,\nparallel}$, K. Banerjee$^1$, P. Roy$^1$, R. Pandey$^1$, G. Mukherjee$^1$, T. K. Ghosh$^1$, J. K. Meena$^1$, T. Roy$^1$, A. Chaudhuri$^1$, M. Sinha$^1$, A. K. Saha$^1$, Md. A. Asgar$^1$, A. Dey$^1$, Subinit Roy$^2$, Md. M. Shaikh$^2$}
\affiliation{$^1$Variable Energy Cyclotron Centre, 1/AF, Bidhan Nagar, Kolkata - 700064, India.} 
\affiliation{$^2$Saha Institute of Nuclear Physics,1/AF, Bidhan Nagar, Kolkata 700064, India.}
\date{\today}

\begin{abstract}
 The even-even nucleus $^{26}$Mg has been studied through the reaction $^{27}$Al(d,$^{3}$He) at 25 MeV beam energy. The spectroscopic factors have been extracted upto 7.50 MeV excitation energy using local, zero range distorted wave Born approximation. The comparison of the spectroscopic factors have been done with previously reported values using the same reaction probe. The extracted spectroscopic factors for different excited states were found to be in good agreement with the previously reported values for the same. The present results were also compared with the predictions from shell model as well as rotational model. The analog states of $^{26}$Al and $^{26}$Mg were found to be in good agreement.
\end{abstract}


\maketitle

\section{Introduction}
The single nucleon pick up reactions like (p,d),(d,t) and (d,$^{3}$He), the stripping reactions like (d,p) are the powerful tools to determine the excitation energy, spin, parity, orbital and total angular momentum and also the spectroscopic factors for single particle levels of the nuclei of interest. The study of the nucleus $^{26}$Mg is important in nuclear physics as well as from nuclear astrophysics point of view as it is the radioactive decay product of $^{26}$Al.~The nucleus $^{26}$Al is the first cosmic radioactivity detected through its characteristic $\gamma$ rays in the interstellar medium and its importance and origin have been investigated widely in the previous years~\cite{ref1,ref2,ref3,ref4,ref5,ref6}. Very recent study of the reaction $^{23}$Na($\alpha$,p)$^{26}$Mg stated that the reaction $^{23}$Na($\alpha$,p)$^{26}$Mg directly influences the production of $^{26}$Al~\cite{Almaraz}. As being the radioactive decay product of $^{26}$Al, the observed excess of $^{26}$Mg in meteorites~\cite{Pherson}, presolar dust~\cite{Hoppe, Huss}, the presence of excess $^{26}$Mg resulting from the decay of $^{26}$Al in calcium-aluminium-rich inclusions (CAIs) and ferromagnesian silicate spherules (chondrules) from Allende meteorite~\cite{Bizzarro} shows that the study of $^{26}$Mg is also important to understand the origin of $^{26}$Al. 

In previous years, several particle transfer reactions were performed to study the different excited states of $^{26}$Mg.~Earlier, the reaction $^{27}$Al(d,$^{3}$He) has been studied at 29 MeV~\cite{Vernotte}, 34.5 MeV~\cite{Wildenthal}, 52 MeV~\cite{Wagner} and at 80 MeV~\cite{Arditi}.~Apart from (d,$^{3}$He) reaction channel, other reaction channels such as $^{25}$Mg($\alpha$,$^{3}$He)~\cite{JJ, yassue} and $^{25}$Mg(d,p)~\cite{Burlein} were also used to study $^{26}$Mg. The spectroscopic factors for the excited states of $^{26}$Mg have been extracted in these reactions. A compilation of different excited states of $^{26}$Mg has been performed by Endt $et$ $al.$~\cite{Endt, Endt1}. 

The Primary motivation of the present study was to examine the dependence of spectroscopic factors on optical model potential parameters used in the analysis by comparing the present results with previously reported values for the same reaction probe. Nine excited states of $^{26}$Mg were studied by zero range distorted wave Born approximation~(ZRDWBA) calculations performed using DWUCK4 code~\cite{kunz} and the extracted values of spectroscopic factors were compared with the previous results and also with predictions from the shell model given in~\cite{Wildenthal,Wagner} and predictions given from rotational model in~\cite{Wagner}. The spectroscopic factors of T=1 analog states of $^{26}$Al and $^{26}$Mg should be identical, so the verification of T=1 analog states in $^{26}$Al and $^{26}$Mg produced in (d - t/~$^{3}$He) reaction sequence was also the motivation of the present study, which is the first study using $^{27}$Al(d, t) and $^{27}$Al(d,$^{3}$He) reactions sequence.   

\section{Experimental Details}
\begin{figure*} [ht]
\includegraphics[width=14cm,height=7cm,clip]{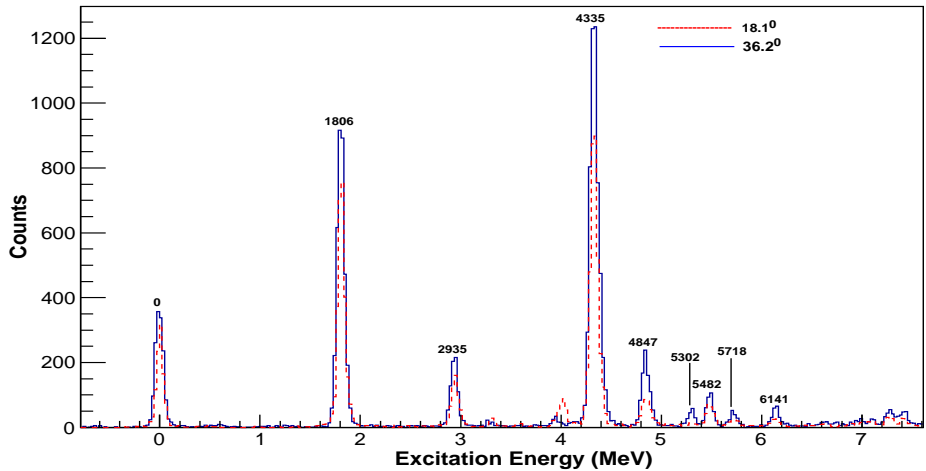}
\caption{\label{fig1}
(Color online)Excitation energy spectra of $^{26}$Mg obtained in the reaction d(25 MeV) + $^{27}$Al at $\theta_ {lab}$ $\approx$ 18$^\circ$ (dash line) and at $\theta_ {lab}$ $\approx$ 32$^\circ$ (solid line).}
\end{figure*}

The experiment was performed using 25 MeV deuteron beam from the K130 Cyclotron on a self-supported $^{27}$Al (thickness$\sim$90 $\mu$g/cm$^2$) target at the Variable Energy Cyclotron Centre, Kolkata.~The experimental details and a sample of typical two dimensional spectrum of light charged particles obtained using Si(E) - Si($\Delta$E) telescope set up have been given in~\cite{vishal}.~A typical excitation energy spectrum of $^{26}$Mg, obtained from the $^{3}$He ridge, which was populated via the reaction channel $^{27}$Al(d,$^{3}$He), is shown in Fig.~\ref{fig1}.~Calibration of detectors for $^{26}$Mg has been done with ground state of $^{26}$Al from the $^{3}$H and three states of $^{26}$Mg (0, 1808 and 2938 keV) from the $^{3}$He spectrum. The systematic and statistical errors have been taken to estimate total error in experimental data. The energy loss corrections due to target thickness and the dead layers in Si detectors have been taken into consideration to identify the excitation energies.~The estimated uncertainties in excitation energies are $\approx$~$\pm 13$ keV.

\section{DWBA Analysis}
\subsection{Optical Model Potential(OMP) Parameters}
The OMP parameters for entrance channel were extracted using the optical model code ECIS94~\cite{ranyal} and detailed description about the extraction of OMP parameters has been given in our previous study of $^{27}$Al(d,t)~\cite{vishal}. We checked all of the three sets of OMP parameters given in~\cite{vishal} as the entrance channel OMP parameters to study the present $^{27}$Al(d,$^{3}$He) reaction at 25 MeV. However, in the present work only one set of OMP parameters from Ref.~\cite{vishal}(set A in Table~\ref{table1}) had been used for the entrance channel.~We used two sets of the OMP parameters for $^{3}$He+$^{26}$Mg partition in the exit channel which are given in Table~\ref{table1} as set B and set C.~The OMP parameter set B was calculated using the relation given in Ref.~\cite{perey} while the set C of OMP parameters was taken from the~\cite{Vernotte}.~The t+$^{26}$Al parameter set given in Ref.~\cite{vishal}, which was calculated using the relation given in~\cite{perey}; was also tested in the present analysis for $^{3}$He+$^{26}$Mg and this parameter set was also found to reproduce the experimental data nicely.~In this paper, we used the sets B and C as given in Table~\ref{table1} as the exit channel potential parameter for the analysis of the reaction $^{27}$Al(d,$^{3}$He). 
\begin{table*}[t]
\caption{The best fit potential parameters used in DWUCK4 code for $^{27}$Al(d,$^{3}$He) reaction.}
\begin{tabular}{ccccccccccccc}\hline\hline
Reaction&Set&V&$R_{o}$&$a_{o}$&$W_{v}$&$W_{D}$&$R_{I}$&$a_{I}$&$V_{ls}$&$R_{ls}$&$a_{ls}$&$R_{C}$ \\
{}&{}&($MeV$)& ($fm$)& ($fm$)& ($MeV$)& ($MeV$)& ($fm$)& ($fm$)&($MeV$)&( $fm$)& ($fm$) & ($fm$)\\
\hline

d+$^{27}$Al         & {A} &  89.209   & 1.061    & 0.701      &  {}     & 2.250    & 1.360  & 0.850  & 9.00  & 1.061   & 0.801 & 1.25\\

$^{3}$He+$^{26}$Mg  & {B} &  151.97   & 1.20     & 0.720      &  37.75  & {}       & 1.400  & 0.880  & 2.50  & 1.20    & 0.720 & 1.30\\

                    & {C} &  217.6    & 1.15     & 0.636      &  32.5   & {}       & 1.319  & 0.986  & {}    & {}      & {}    & 1.40\\
                    
p+$^{26}$Mg         & {} &   D        & 1.20    &  0.650      &  {}    & {}        & {}     & {}     & {}    & {}      & {}    & 1.25\\

\hline 
\multicolumn{13}{l}{Set A Parameters were taken from the Ref.~\cite{vishal}.}\\
\multicolumn{13}{l}{Set B were extracted from the relation given in Perey and Perey~\cite{perey} .}\\
\multicolumn{13}{l}{Set C Parameter sets were taken from $^{27}$Al(d,$^{3}$He)~\cite{Vernotte}.}\\
\multicolumn{13}{l}{D, the Well depth adjusted to give the required separation energy for the transferred particle.}
\end{tabular}
\label{table1}
\end{table*}

\subsection{Calculation of ${C^{2}S}$ and uncertainty in ${C^{2}S}$ values }
The experimental angular distributions of cross sections of the observed excited states of $^{26}$Mg upto 7.50 MeV are shown in Figs.~\ref{fig2}~and~\ref{fig3} and these were fitted with theoretical predictions in ZRDWBA using the computer code DWUCK4~\cite{kunz}.~The theoretical predictions were found to reproduce the experimental data well for lower excited states but for higher excited states the deviations between theoretical predictions and experimental data were found to be large at backward angles.~The value of transferred angular momentum was calculated using the prescription given in~\cite{sach}.~Spectroscopic factors of these states of $^{26}$Mg were extracted using the relation between experimental cross section and theoretical cross sections given in~\cite{bassel};

\begin{equation}
\mathrm{\left(\frac{d\sigma}{d\Omega}\right)_{exp.}}=2.95\frac{C^{2}S}{2J+1}\mathrm{\left(\frac{d\sigma}{d\Omega}\right)_{DWUCK4}},
\end{equation}
where, $\mathrm{(\frac{d\sigma}{d\Omega})_{exp.}}$ is the experimental
differential cross-section and $\mathrm{(\frac{d\sigma}{d\Omega})_{DWUCK4}}$ is
the cross-section predicted by the DWUCK4 code. $J$ ($J =\it l \pm \frac{1}{2}$) is the total angular momentum of the orbital from where proton is picked up. $C^{2}$is the isospin Clebsch-Gordon coefficient and its values for the reactions $^{27}$Al(d,t)$^{26}$Al and $^{27}$Al(d,$^{3}$He)$^{26}$Mg are 1/3 and 2/3, respectively, and  S is the spectroscopic factor.\\

To extract ${C^{2}S}$ values for different excited states of $^{26}$Mg, we used two combinations A-B and A-C of OMP parameters obtained from the entrance-exit channel potential parameter sets given in Table~\ref{table1}.~The variation in the extracted spectroscopic factors between the potential combinations A-B and A-C, was less than 25\% for $\it l= 2$ transfer (It should be noted that the variation includes all the observed states for which configuration mixing ${0d}_{5/2}$ + ${1s}_{1/2}$ was considered and also for those states which were analysed without configuration mixing).~This variation for the $\it l= 0$ transfer for the states shown in Fig.~\ref{fig2} is large (maximum upto 67\%) because of very low contribution as compared to $\it l= 2$ transfer while for the states shown in Fig.~\ref{fig3}, the variation was found to be from 14\% to 30\%.~The average of the extracted ${C^{2}S}$ values for different excited states of $^{26}$Mg from the said two combinations was taken as final and are given in Table~\ref{table2}. The estimated deviations between the ${C^{2}S}$ values extracted individually from the combinations A-B and A-C were used in determining the uncertainties in the ${C^{2}S}$ values for different excited states of $^{26}$Mg.~The theoretically predicted ${C^{2}S}$ values from the shell and rotational models given in ~\cite{Wagner, Arditi} are also given in Table~\ref{table2} for comparison. 
\begin{figure}
\centering
\includegraphics[width=9cm,height=13cm,clip]{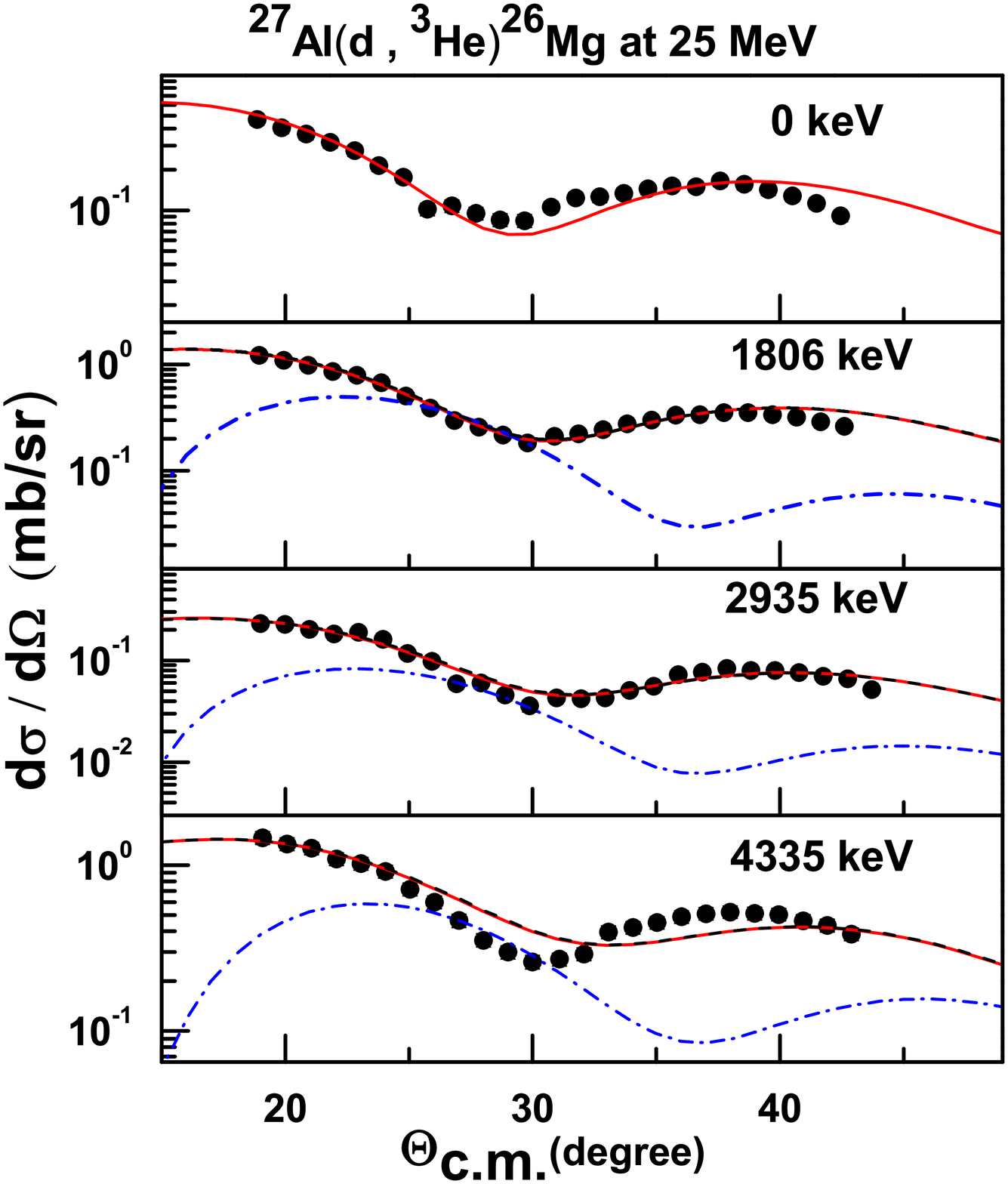}
\vspace{-1.5cm}
\caption {\label{fig2}
(Color Online) Angular distributions for ground, 1806, 2935 and 4335 keV states.The filled circles represent experimental data points and solid line represents theoretical cross section from DWUCK4 code for the OMP combination A-B for considering pick up from ${0d}_{5/2}$ only, dash-dot-dash represents theoretical cross section for A-B with pick up from ${1s}_{1/2}$ only while dash-dash represents the theoretical cross sections for ${0d}_{5/2}$ + ${1s}_{1/2}$ mixing. Note that theoretical cross sections were normalized to experimental data points.}
\label{fig2}
\end{figure}%
\begin{figure}
\centering
\includegraphics[width=9cm,height=13cm,clip]{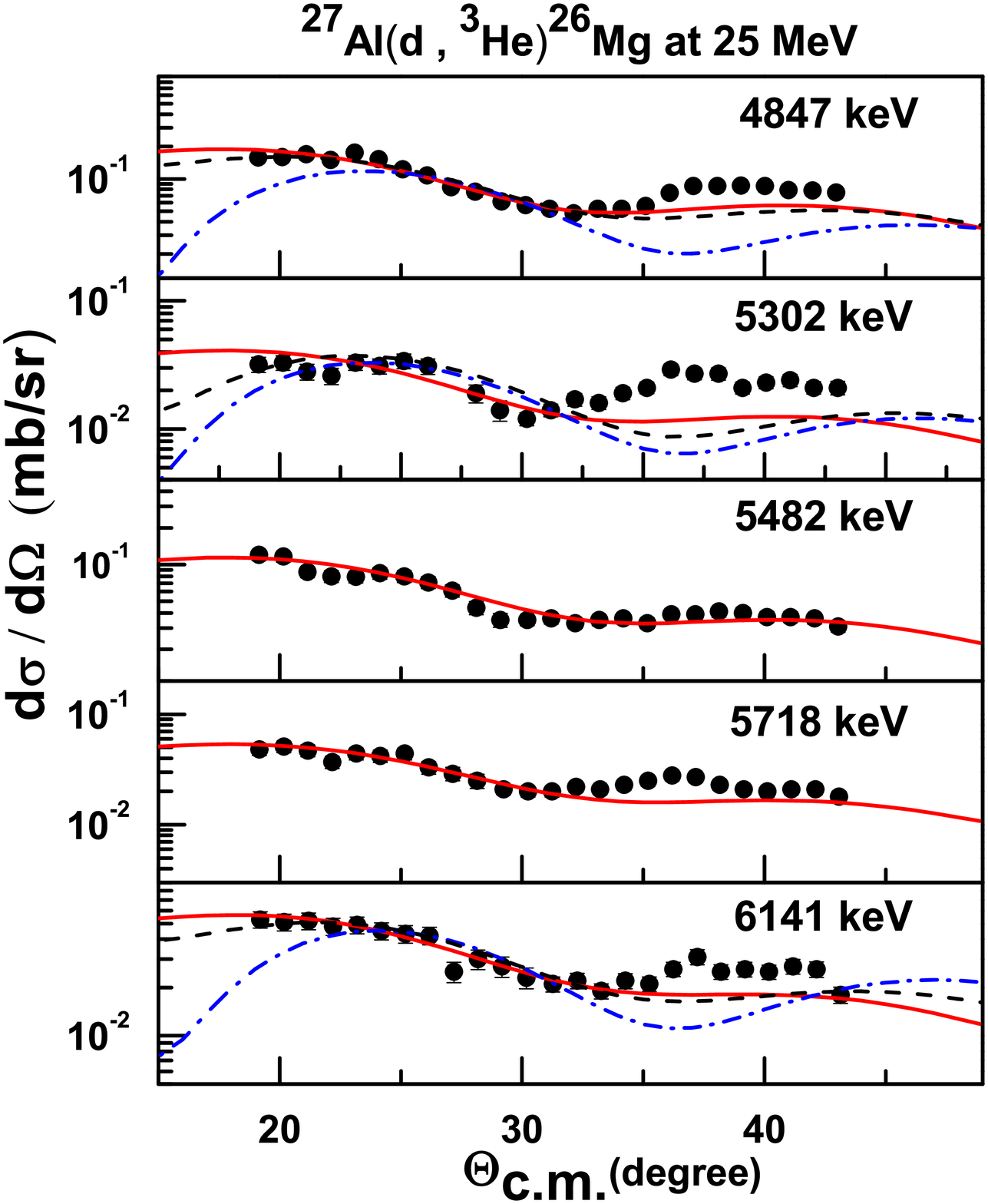}
\caption {\label{fig3}
(Color Online) Angular distributions for 4847, 5302, 5482, 5718 and 6141 keV states (same notations as in Fig.\ref{fig2}).} 
\end{figure}

\begin{table*}[t]
\centering
\caption{Extracted values of ${C^{2}S}$ for different excited states of $^{26}$Mg from the reaction $^{27}$Al(d,$^{3}$He) at 25 MeV.}
\begin{tabular}{cccccc ccccc ccc ccc} \hline\hline
{}&{}&{}&{}&{}&{}&{}&{}&{}&{}&{}&{}&{}&{}&{}&{}&{}\\
{$^{a)}$Ex}&{}&{J$^{\pi}$}&{}&{$^{b)}$Ex}&{}&{}&{}&{$^{b)}{C^{2}S}$}&{}&{}&{}&{$^{\star}$TC1}&{}&{}&{$^{\dagger}$TC2}&{}\\
{keV}&{}&{}&{}&{keV}&{}&{}&{l=0}&{}&{l=2}&{}&{l=0}&{}&{l=2}&{l=0}&{}&{l=2}\\
{}&{}&{}&{}&{}&{}&{}&{}&{}&{}&{}&{}&{}&{}&{}&{}&{}\\
\hline
{}&{}&{}&{}&{}&{}&{}&{}&{}&{}&{}&{}&{}&{}&{}&{}&{}\\

{0}&{}&{0$^{+}$}&{}&{0}&{}&{}&{}&{}&{0.17$\pm0.05$}&{}&{...}&{}&{0.29}                            &{...}&{}&{0.33}\\
{1808.7}&{}&{2$^{+}$}&{}&{1806}&{}&{}&{0.002}&{}&{0.57$\pm0.14$}&{}&{0.014}&{}&{0.75}&{0}&{}&{0.60}\\
{2938.3}&{}&{2$^{+}$}&{}&{2935}&{}&{}&{0.002}&{}&{0.13$\pm0.03$}&{}&{0.0032}&{}&{0.29}&{...}&{}&{...}\\
\Bigg\{{\begin{tabular}{c}
 4318\\
 4332\\
 4350\\
 \end{tabular}
}&{}&{(2,4)$^{+}$}&{}&{4335}&{}&{}&{0.004}&{}&{1.03$\pm0.19$}&{}&\Bigg\{{\begin{tabular}{c}
 ...\\
 ...\\
 0.16\\
 \end{tabular}}&{}&\Bigg\{{\begin{tabular}{c}
 ...\\
 1.80\\
 0.002\\
 \end{tabular}}                    &\Bigg\{{\begin{tabular}{c}
                                                                                      ...\\{$\approx 0$}\\{$\approx 0$}\\ 
                                                                                        \end{tabular}}
                                                                                        &{}     &\Bigg\{{\begin{tabular}{c}
                                                                                        0.07\\ 0.50\\ 0.14\\ \end{tabular}}\\
{4835.1}&{}&{2$^{+}$}&{}&{4847}&{}&{}&{$^{\ddagger}$0.011$\pm0.010$}&{}&{0.11$\pm0.02$}&{}            &{0.0061}&{}&{0.022}      &{$\approx 0$}&{}&{0.36}\\
{5291.7}&{}&{2$^{+}$}&{}&{5302}&{}&{}&{0.011$\pm0.004$}&{}&{0.011$\pm0.004$}&{}            &{...}&{}&{0.018}        &{...}&{}&{...}\\
{5476.1}&{}&{4$^{+}$}&{}&{5482}&{}&{}&{}&{}&{0.112$\pm0.024$}&{}            &{...}&{}&{0.025}     &{...}&{}&{...}\\
{5715.6}&{}&{4$^{+}$}&{}&{5718}&{}&{}&{}&{}&{0.05$\pm0.006$}&{}            &{...}&{}&{...}    &{...}&{}&{...}\\
{6125.5}&{}&{3$^{+}$}&{}&{6141}&{}&{}&{$^{\ddagger}$0.011$\pm0.007$}&{}&{0.043$\pm0.005$}&{}            &{...}&{}&{...}     &{...}&{}&{...}\\
{}&{}&{}&{}&{}&{}&{}&{}&{}&{}&{}&{}&{}&{}&{}&{}&{}\\
\hline\hline
\multicolumn{17}{l}{$^{a)}$ Values taken from the NNDC~\cite{nndc},        $^{b)}$ Present work.}\\
\multicolumn{17}{l}{$^{\star}$TC1 represents theoretical predictions from shell model given in~\cite{Wildenthal,Wagner}. }\\
\multicolumn{17}{l}{$^{\dagger}$TC2 represents theoretical predictions from rotational model given in~\cite{Wagner}.}\\
\multicolumn{17}{l}{$^{\ddagger}$Due to uncertainty in data, the reported uncertainties are greater than 50\%.}\\

\end{tabular}
\label{table2}
\end{table*}

\begin{table}[t]
\centering
\caption{Comparison of ${C^{2}S}/{C^{2}S}_{gs}$ obtained from different reactions for different excited states of $^{26}$Mg. }
\begin{tabular}{ccccccc}\hline\hline
{}&{}&{}&{}&{}&{}&{} \\

$^{a)}$Ex&\it l&(d,$^{3}$He)$^{a}$&(d,$^{3}$He)$^{12}$&(d,$^{3}$He)$^{13}$&(d,$^{3}$He)$^{14}$&(d,$^{3}$He)$^{15}$ \\
(keV)&{}&{}&{}&{}&{}&{} \\
{}&{}&{}&{}&{}&{}&{} \\

\hline
0               & 2            & 1             &  1             &  1             &  1            &  1   \\       
1806            & 2            & 3.35          &  3.33          &  3.56          &  3.44         &  3.34\\    
2935            & 2            & 0.76          &  0.73          &  0.77          &  0.70         &  1.11  \\      
4335            & 2            & 6.06          &  {}            &  7.10          &  7.15         &  8.19\\    

4847            & 2            & 0.65          &  0.27          &  $<$0.53                  &  {} \\               

5302            & 2            & 0.06          &  0.04          &  {}            &  {}           &  {}  \\    
5482            & 2            & 0.66          &  0.70          &  0.80          &  1.19         &  0.96  \\      
5718            & 2            & 0.29          &  0.23          &  {}            &  {}           &  {}\\    
6141            & 2            & 0.25          &  0.24          &  0.4           &  {}           &  {} \\      
\hline\hline
\multicolumn{7}{l}{$^a$ Represents the values listed in Table II (Present work).}\\
\end{tabular}
\label{table3}
\end{table}
The uncertainties in the ${C^{2}S}$ values that arise due to the choice of bound-state potential parameters in the calculation of ZRDWBA, were also taken into consideration to determine the uncertainties in ${C^{2}S}$ values in the present study.~To see the effect of radius of bound-state potential, we extracted ${C^{2}S}$ values for the two afore said OMP combinations using $R_{o}$=1.25 $fm$ and taken the average of the ${C^{2}S}$ values for all the states.~The difference between the average ${C^{2}S}$ values calculated using the radius of bound-state potential $R_{o}$=1.25 $fm$ and average using $R_{o}$=1.20 $fm$ as described above keeping other parameters unchanged was also included in the uncertainty in ${C^{2}S}$ values of the observed states of $^{26}$Mg.~The reduction in the ${C^{2}S}$ values for $R_{o}$=1.25 $fm$ was approximately 5 to 25\% as compared with $R_{o}$=1.20 $fm$ for $\it l= 2$ transfer while for $\it l= 0$ transfer, 15\% to 40\% reduction in ${C^{2}S}$ values were noticed except for the states at 1806 and 2935 keV(For these two states, it was found to be increased by 50\%).

Along with the above two types of differences estimated in the calculation of ${C^{2}S}$ values, the average error in experimental data points was also included in the uncertainty of ${C^{2}S}$ values for different excited states of $^{26}$Mg.~The total uncertainty calculated has been listed in Table~\ref{table2} along with the measured ${C^{2}S}$ values for different excited states of $^{26}$Mg.~So, to reduce the effect in absolute normalization due to the choice of optical model potential parameters and other parameters that can affect the ${C^{2}S}$ value, we calculated the relative spectroscopic factors (keeping ground state spectroscopic factor to be 1) for different excited states of $^{26}$Mg and compared our results with previously reported values for the same reaction. The relative spectroscopic factors are listed in Table III. 

It was clear from Table~\ref{table2} that the ${C^{2}S}$ values extracted for the states of $^{26}$Mg shown in Fig.~\ref{fig2} were very large for $\it l= 2$ transfer as compared with $\it l=0$ transfer in ${0d}_{5/2}$ + ${1s}_{1/2}$ configuration mixing. So, for $\it l=0$ transfer, the uncertainties were not shown for the states shown in Fig.~\ref{fig2} and also we may avoid configuration mixing for these states but for the sake of completeness we presented the results assuming configuration mixing.~For the states shown in Fig.~\ref{fig3}, the contribution of ${C^{2}S}$ values extracted for $\it l= 0$ transfer has increased as compared with the states shown in Fig.~\ref{fig2}.

\subsection{Discussion of extracted ${C^{2}S}$ factors.}

The fitted experimental cross sections for ground, 1806, 2935 and 4335 keV states alongwith their respective theoretical cross sections predicted from DWUCK4 code are shown in Fig~\ref{fig2}. The extracted ${C^{2}S}$ value for ground state was found to be less in the present study; however, it is consistent, within the limits of uncertainties, with previously reported values using same reaction probe at different energies. ${C^{2}S}$ value of the 1806 keV state for $0d_ {5/2}$ in the present analysis was also found to be less compared with the previously reported experimental values given in ~\cite{Vernotte, Wildenthal, Wagner, Arditi} but the relative ${C^{2}S}$ value was found to be in agreement with previously reported values. It is also found to be in agreement with the rotational model predictions given in ~\cite{Wagner}. The ${C^{2}S}$ value in the present analysis for $1s_ {1/2}$ mixing was found to be in good agreement with both experimentally reported value given in ~\cite{Vernotte} as well as predictions from shell and rotational models given in ~\cite{Wagner}. Though the extracted ${C^{2}S}$ values for ground, 1806 keV and 2935 keV excited states were found to be less than the previously reported values for the same however relative ${C^{2}S}$ values were found to be consistent with previously reported values for these states.~The ${C^{2}S}$ values for 2935 keV state for both $\it l= 2$ and $\it l= 0$ transfers were also less compared with previously reported values; however, the  relative ${C^{2}S}$ value for $\it l= 2$ transfer is in agreement with the previously reported values while that for $\it l= 0$ transfer is comparable with the value reported in ~\cite{Wagner}. Around the excited state at 4335 keV, there are 4318 keV, 4332 keV and 4350 keV excited states and in the present study these states could not be resolved. So the analysis was performed by taking the centroid position and spectroscopic factor for the 4335 keV excited state was extracted.~The spectroscopic factor for the 4335 keV state was found to be less in the present study.
 
In Fig.~\ref{fig3}, we have fitted experimental cross sections for 4847, 5302, 5482, 5718 and 6141 keV states with the respective theoretical cross sections predicted from DWUCK4 code. The extracted ${C^{2}S}$ value for the state at 5302 keV was found to be in good agreement with previously reported experimental value in ~\cite{Vernotte} for both $\it l= 2$ and $\it l= 0$ transfers and also in good agreement with theoretical predictions from shell model given in ~\cite{Wildenthal, Wagner}. The extracted ${C^{2}S}$ value for the 5482 keV state was also less in the present study and its relative ${C^{2}S}$ value was compared with previously reported experimental values as well as with predictions from shell model given in ~\cite{Wildenthal, Wagner}. It was found that its relative ${C^{2}S}$ is in agreement with the values reported in Ref.~\cite{Vernotte, Wildenthal}.~The extracted ${C^{2}S}$ value for the 6141 keV state was found to be in agreement with the previously reported value in ~\cite{Vernotte} for $\it l= 2$ and $\it l= 0$ transfers. The relative ${C^{2}S}$ value of 6141 keV state was found to be in good agreement with previously reported values listed in Table III. There may be mixture of two or three states in 4847 keV state but we have taken them together as a single state and the reported excitation energy corresponds to the centroid position. Similarly, there may be mixture of two states in the state at 5718 keV. For 4847 and 5718 keV states the extracted spectroscopic factors were compared with the spectroscopic factors reported earlier for those states which are close to the two states and are listed in Table II and III.  
\subsection{Analog states of $^{26}$Al and $^{26}$Mg.}
For T=1 states in $^{26}$Al, the spectroscopic factors should be identical to their  $^{26}$Mg analogs. So, we compared spectroscopic factors of analog states of $^{26}$Mg and $^{26}$Al, extracted from the present study of (d,$^{3}$He) reaction and our previously studied (d,t) reaction~\cite{vishal}. The comparison is given in Table~\ref{table4}. Although the ${C^{2}S}$ values of $^{26}$Al and $^{26}$Mg are not same but from Table~\ref{table4}, it is clear that the ratio of S values for $^{26}$Al and $^{26}$Mg are consistent within 11\% and it can be take care by the uncertainties in the ${C^{2}S}$ values of the respective states. So, one can draw the conclusion from Table~\ref{table4} that the analog states are appreciably excited.

\begin{table}[t]
\centering
\caption{Comparison of analog states from the reactions $^{27}$Al(d,t)$^{26}$Al and $^{27}$Al(d,$^{3}$He)$^{26}$Mg at 25 MeV for T=1.}
\begin{tabular}{cccccc} \hline\hline
{}&\hspace{0.4cm}{}&\hspace{0.4cm}{}&\hspace{0.4cm}{}&\hspace{0.4cm}{}&\hspace{0.4cm}{}\\

$^{a)}$Ex     &\hspace{0.4cm}$^{(a)}$${C^{2}S}$   &\hspace{0.4cm}J$^{\pi}$          &\hspace{0.4cm}$^{b)}$Ex         & \hspace{0.4cm}$^{(b)}{C^{2}S}$  &\hspace{0.4cm}$S\left(=\frac{a}{b}\right)$         \\
(keV)&\hspace{0.4cm}{}&\hspace{0.4cm}{}&\hspace{0.4cm}{}&\hspace{0.4cm}{}(keV)&\hspace{0.4cm}{}\\
{}&\hspace{0.4cm}{}&\hspace{0.4cm}{}&\hspace{0.4cm}{}&\hspace{0.4cm}{}&\hspace{0.4cm}{}\\

\hline 


230           & \hspace{0.4cm}0.09$\pm0.03$           &\hspace{0.4cm} 0$^{+}$                 &\hspace{0.4cm}  0               & \hspace{0.4cm} 0.17$\pm0.06$   &\hspace{0.4cm}  1.04\\ 

2070          & \hspace{0.4cm}0.26$\pm0.06$           &\hspace{0.4cm} 2$^{+}$                 &\hspace{0.4cm}  1806            & \hspace{0.4cm} 0.57$\pm0.13$   &\hspace{0.4cm}  0.91\\ 

3160          & \hspace{0.4cm}0.06$\pm0.01$           &\hspace{0.4cm} 2$^{+}$                 & \hspace{0.4cm} 2935            & \hspace{0.4cm} 0.13$\pm0.04$   & \hspace{0.4cm} 0.92\\ 
\hline\hline 
\multicolumn{6}{l}{$S$ is the spectroscopic factor only without $C^{2}$.}\\
\multicolumn{6}{l}{$^a$ Values taken from the study of the reaction $^{27}$Al(d,t)~\cite{vishal}.}\\
\multicolumn{6}{l}{$^b$ Present study of the reaction $^{27}$Al(d,$^{3}$He).}\\
\end{tabular}
\label{table4}
\end{table}

\section{Summary and Conclusion}\label{Summary and Conclusion}
Different excited states of $^{26}$Mg have been populated through the reaction $^{27}$Al(d, $^{3}$He)$^{26}$Mg at 25 MeV beam energy and studied with zero range distorted wave Born approximation. Comparison of the measured spectroscopic factors with those reported earlier for the same reaction probe has been found to be in good agreement. Extracted ${C^{2}S}$ values for the analog states of $^{26}$Al and $^{26}$Mg were compared and were found to be in good agreement indicating that the ratios of the production of analogs were equally probable.

The potential parameter dependence of spectroscopic factors had also been checked with two different exit channel potential parameters. In the present study of the reaction $^{27}$Al(d,$^{3}$He), less than 25\% variation has been observed in extracted ${C^{2}S}$ values of the states of $^{26}$Mg for $\it l= 2$ transfer between the two combination of entrance-exit channel potential parameters while for $\it l= 0$ contribution, the variation was in the range between 14 to 67\%. It has also been verified that the change in the radius of bound-state potential affected significantly and reduced the extracted ${C^{2}S}$ values up to 25\% for $\it l= 2$ transfer while for $\it l= 0$ transfer, it has decreased upto a maximum of 40\%. All those factors that affected ${C^{2}S}$ values, which have been discussed in the present paper, were used to estimate the uncertainties in the extracted ${C^{2}S}$ values. 

In conclusion, $^{26}$Mg can be studied well with the reaction $^{27}$Al(d,$^{3}$He).~The experimental data was found to be well reproduced by the theoretical predictions. The extracted values of spectroscopic factors were found to be in good agreement with the predictions from shell model, rotational model and also with those reported earlier.

\section{ACKNOWLEDGEMENT}The authors thank the cyclotron operating staff for their cooperation during the experiments. One of the authors (A.D.) acknowledges with thanks the financial support provided by the Science and Engineering Research Board, Department of Science and Technology, Government of India vide project No. SR/FTP/PS067/2012 dated 10/06/2012. One of the authors (S.B.) acknowledges with thanks the financial support received as Raja Ramanna Fellow from the Department of Atomic Energy, Government of India.

\end{document}